# Liver Lesion Detection with Task-Specific Organ-Level Tube Current Modulation: A Retrospective Reader Study

Maria Jose Medrano[1]*; Sen Wang[1]*; Kaylee W. Fang[1]; Jennie Cao[1]; Liyan Sun[1]; Luyao Shen[1]; Andy Shon[1]; Luke Yoon[1]; Grant M. Stevens[2]; Adam S. Wang[1]; Justin R. Tse[1]

[1]Department of Radiology, Stanford University School of Medicine, Stanford University, Stanford, California, USA

[2]GE HealthCare, Menlo Park, California, USA

*Maria Jose Medrano and Sen Wang contributed equally to this work.

---

**Abstract**

**Rationale and Objectives:** To evaluate whether task-specific organ-level TCM improves hepatic lesion detection and abdominal image quality compared with uniform dose reduction at equivalent effective dose.

**Materials and Methods:** In this single-center, IRB-approved, retrospective study, 22 oncology patients with 68 liver lesions (38 malignant, 30 benign) underwent chest–abdomen–pelvis CT. Using projection-domain noise emulation, five series were generated per patient: original acquisition (Orig-100); uniform reduction to 40% (UD-40) and 60% (UD-60) of effective dose; and organ-level TCM at matched 40% (Opt-40) and 60% (Opt-60). Three abdominal radiologists independently detected and classified lesions as benign or malignant and rated image quality on a 5-point Likert scale. Per-lesion sensitivity and image-quality scores were compared using the McNemar and Wilcoxon signed-rank tests, respectively. Malignant lesion-size thresholds ($x_{50}$, $x_{90}$) were estimated by logistic regression.

**Results:** At 60% dose, Opt-60 malignant sensitivity was comparable to Orig-100 (82% vs 80%) and higher than UD-60 (67%). At 40% dose, Opt-40 exceeded UD-40 (68% vs 52%). All-lesion detection improved similarly (Opt vs UD: 64% vs 54% at 40%; 77% vs 65% at 60%), with Opt-60 comparable to full dose (77% vs 76%). All Opt-versus-UD detection differences were significant ($p \leq 0.003$). Opt-60 lesion-size thresholds ($x_{50}$= 3.3 mm; $x_{90}$= 15.7 mm) matched Orig-100 (3.7 mm, 16.8 mm). Abdominal image quality was higher with Opt than UD at both doses.

**Conclusion:** At equivalent effective dose, organ-level TCM improved liver lesion detection over uniform reduction, with Opt-60 comparable to full dose. These findings support organ-level optimization for dose reduction while preferentially preserving image quality for liver metastasis detection.

## 1. Introduction

The liver represents one of the most common sites of metastases across all cancers.[1–3] Contrast-enhanced CT is the primary imaging modality used for their detection and surveillance.[4] Most hepatic metastases enhance differently relative to surrounding parenchyma, but often the attenuation differences are subtle and challenging to detect on CT.[5,6] Consequently, accurate liver metastasis detection requires sufficient image quality and contrast-to-noise ratio.

Noise can be reduced in CT by increasing radiation dose. However, this represents a trade-off in long-term safety, particularly in oncology patients who are expected to undergo lifelong surveillance CT.[7] Thus, an ongoing goal has been to reduce radiation exposure while preserving diagnostic image quality in liver metastasis detection. Prior strategies have used conventional filtered back projection, combined automatic tube voltage selection and tube current modulation, advanced iterative reconstruction, and denoising techniques.[8–12] However, these approaches primarily quantified dose using scanner-output metrics such as CT dose index volume ($CTDI_{vol}$) and dose-length product (DLP), or effective dose estimated from DLP using population-average conversion factors.[13] Such metrics reflect overall scanner output or population-based estimates rather than accounting for differences in radiation dose across a patient's anatomy. Furthermore, prior organ-level TCM approaches have demonstrated the potential to preserve image quality in selected organs at reduced overall dose,[14–17] but the effect of task-specific dose redistribution on hepatic lesion detection has not been evaluated.

A recently proposed organ-level TCM method enables task-specific optimization of tube current profiles by preferentially preserving image quality in selected organs while maintaining a predefined effective-dose target.[18] The purpose of this study was to evaluate whether task-specific organ-level TCM can improve hepatic lesion detection and abdominal image quality compared with uniform dose reduction at equivalent effective-dose targets.

## 2. Materials and Methods

### *2.1 Patient Selection*

From March 13, 2023 to June 23, 2023, 287 patients at a comprehensive cancer center underwent outpatient CT chest-abdomen-pelvis in single-energy mode. A research coordinator reviewed the electronic medical record (EMR) and excluded patients who did not have a history of cancer (n=11). The research coordinator reviewed the CT reports of patients with a history of cancer to determine if liver metastases were present (n=38). A board certified, fellowship-trained, abdominal radiologist with 4 years of post-training experience subsequently reviewed the original clinical images and the EMR on a clinical workstation to exclude patients based on the following criteria: 1) widespread liver metastases (defined as >7 metastases) and 2) lack of a reference standard, defined as tissue confirmation or subsequent imaging follow-up (MRI, CT, or PET/CT). For patients who underwent multiple CT exams during the time period, the first CT was selected. At this stage, all liver lesions (benign and malignant) were annotated, and the long-axis dimension of each lesion was measured in the axial plane. After applying all exclusion criteria, this resulted in 11 patients with 38 metastases. A subset of control cases with cancer but without liver metastases was selected to match the age, sex, and body-mass index of the final patient cohort. Figure 1 depicts the process to derive the patient cohort.

### *2.2 CT Acquisition*

Patients were imaged on a GE HealthCare Revolution Apex (Waukesha, WI) scanner using a single-energy CT chest-abdomen-pelvis protocol at our institution. Weight-based iodinated contrast (Isovue-370, Bracco Diagnostics Inc., Monroe Township, NJ, USA) was administered intravenously and images were obtained in the portal venous phase. CT exams were acquired in helical mode using the vendor's automatic exposure control (SmartmA, GE HealthCare), which modulates scanner output based on the scout scan and a weight-based noise index defining the target image quality. Table I summarizes acquisition parameters, including $CTDI_{vol}$, effective dose, and noise index. Hereafter, the original clinical acquisition obtained using the scanner's proprietary automatic exposure control is referred to as *Orig-100*. Original reconstructed images, along with projection and tube current modulation data, of the final patient cohort and controlled cases were exported from the clinical CT scanner at our institution.

### *2.3. Image Generation*

#### *Organ-Level Noise and Dose Calculations*

Image noise maps at different tube current modulation (TCM) dose levels were estimated using a previously validated empirical noise model.[19] This noise model uses noise maps derived from difference images at two synthetic dose levels (20% and 90% in this study) to spatially estimate varying image noise as a function of dose while accounting for contributions from reconstruction-related effects, as well as quantum and electronic noise. The model was previously validated using simulation, clinical, and experimental phantom data, demonstrating good agreement between estimated and measured image noise across the dose range relevant to this study.[19]

Patient-specific radiation dose maps were calculated using the publicly available MC-GPU framework, with previously described modifications to incorporate scanner-specific components including the anode heel effect, bowtie filter, and TCM.[19] Patient-specific organ segmentations were generated using the open-source TotalSegmentator v2.5.0 and applied to the corresponding dose and noise maps to obtain organ-level dose and noise estimates.[20] For effective dose calculations, bone marrow and bone surface masks were derived from bone segmentations using morphological erosion (SciPy v1.13.1; square connectivity = 1), with remaining and eroded portions assigned to bone marrow and bone surface, respectively. The resulting patient-specific organ doses were subsequently combined using the tissue-weighting scheme described below to calculate the effective-dose metric used for optimization. Further details regarding the organ-level dose and noise estimation frameworks can be found in Wang et al.[19]

*Task-Specific Organ-Level Tube Current Modulation (retroOpt)*

In the implemented retrospective organ-level tube current modulation method[18], *retroOpt AEC,* a TCM profile $\mathrm{TCM}(c)$ is parameterized as a weighted combination of TCM basis functions $\mathrm{TCM}_i$ using the Discrete Cosine Transform (DCT). The optimization determines the corresponding basis coefficients $c_i$ that minimize a weighted summation of estimated organ-level image noise while satisfying a predefined weighted organ-dose constraint (Equation 1).

| | |
|---|---|
| $$\min_{c} \sum_{L} w_L^n N_L(c)$$ | |
| $$s.t. \quad \sum_{L} w_L^d D_L(c) = f \sum_{L} w_L^d D_L(\mathrm{TCM}_0)$$ | Equation 1 |
| $$\mathrm{TCM}(c) = \sum_i c_i \,\mathrm{TCM}_i\,, \;\; \mathrm{TCM}(c) \leq \mathrm{TCM}_0, \;\; \mathrm{mA}_{\min} \leq \mathrm{TCM}(c) \leq \mathrm{mA}_{\max}$$ | |

In Equation 1, $N_L$ and $D_L$ represent the estimated noise and dose, respectively, for organ $L$. The noise weighting factors $w_L^n$, determine the relative image quality emphasis assigned to individual organs, whereas the dose weighting factors $w_L^d$, determine the contribution of individual organ doses to the overall dose constraint. Because organ dose varies linearly with tube current, organ-level doses were calculated for each TCM basis function and subsequently combined according to the corresponding basis coefficients to obtain organ doses for the resulting TCM profile. This allowed organ dose to be evaluated throughout the optimization without repeating the Monte Carlo dose calculation for each candidate TCM profile. Organ-level noise for each candidate TCM profile was estimated using the empirical noise model described in the preceding section. This framework therefore allows image quality to be preferentially preserved in selected organs while maintaining a specified overall dose constraint (Supplementary Materials, Figure S1). The optimization was additionally constrained such that the resulting TCM profile did not exceed the patient's original TCM profile ($\mathrm{TCM_0}$) and remained within the minimum and maximum tube current limits of the scanner ($\mathrm{mA_{min}}$ and $\mathrm{mA_{max}}$, respectively). The former is due to the retrospective nature of our study – we can only emulate scans with tube current less than or equal that of the original scan for all projections.

For the present study, an ICRP Publication 103-based tissue-weighting scheme[21] was used to define the dose weighting factors $w_L^d$ such that the weighted summation of organ doses represented the effective-dose metric used for optimization. Because breast and gonadal tissues were not included in the TotalSegmentator organ set used in this study, their corresponding tissue weighting factors were redistributed to other tissue categories for the effective-dose calculation. Specifically, the gonadal tissue weighting factor (0.08) was reassigned to the bladder, and the breast tissue weighting factor (0.12) was reassigned to the remainder-tissue category, resulting in adjusted weighting factors of 0.12 for the bladder and 0.24 for the remainder tissues. This modified ICRP 103-based weighting scheme was applied consistently across all dose conditions. The dose constraint was set such that the optimized TCM profiles corresponded to 40% or 60% of the effective-dose metric calculated for the patient's original acquisition, resulting in *Opt-40* and *Opt-60*, respectively. Separately, the organ-level noise weighting factors $w_L^n$ were adjusted such that the resulting liver dose remained within dose ranges previously associated with adequate hepatic lesion detection in the literature.[6,10,22,23] RetroOpt TCM optimization is further described in Wang et al.[18]

*Image Generation at Different Effective Dose Levels*

Low-dose images corresponding to Opt-40 and Opt-60 were subsequently reconstructed using a projection-domain noise emulation method that inserts noise to account for quantum and electronic noise at the respective reduced dose levels, followed by image reconstruction.[24] All reconstructions were performed by filtered backprojection with a Standard kernel. To evaluate the effect of task-specific organ-level dose redistribution relative to uniform dose reduction, we generated additional images at the same 40% and 60% patient-specific effective-dose targets using the original vendor TCM profile uniformly scaled in tube current (*UD-40* and *UD-60*, respectively). This overall pipeline resulted in five reconstructions per patient for comparative analysis: *Orig-100, Opt-40, Opt-60, UD-40, and UD-60.*

### *2.4. Image Evaluation*

Three board-certified, fellowship-trained abdominal radiologists with 7, 8, and 17 years of post-training experience independently reviewed CT images across different dose levels and TCM schemes in 5 sessions. A randomized, blinded reading schedule was generated for each reader to evaluate one exam corresponding to UD-40, UD-60, Opt-40, Opt-60, or Orig-100 per session for each patient. Each session was separated by a minimum 30-day wash-out period to reduce recall bias. The imaging consisted of axial images at 2.5 mm section thickness and reformations in the coronal and sagittal plane at 2 mm section thickness, respectively. All imaging was evaluated on a research instance of our institutional PACS. Images were anonymized, and radiologists were blinded to dose level and patient medical records, including histopathology reports, comparison examinations, and if the patients had a history of cancer.

Radiologists were instructed to annotate all liver lesions in the axial plane, classify them as malignant or benign, and measure size. There was no minimum or maximum allowable lesion size. Readers were also asked to independently assign image quality scores for the abdomen (the primary region of interest) and the pelvis (a secondary, non-target region) using a predefined 5-point Likert scale (1 = non-diagnostic image quality, 2 = sub-optimal diagnosis with compromised diagnosis confidence, 3 = acceptable diagnostic image quality, 4 = good diagnostic image quality, 5= excellent diagnostic image quality). Supplementary material S2 provides further details on the instructions followed by each radiologist to define image quality based on the defined Likert scale.

### *2.5. Statistical Analysis*

The size and location of each lesion detected by the readers were recorded for each patient across all dose levels and TCM schemes. Per-reader sensitivity for malignant lesions and for all lesions (malignant and benign combined) was computed on a per-lesion basis by assigning a binary detection outcome (1 = detected, 0 = not detected) relative to a reference standard established from prior imaging and patient medical records. A liver lesion was considered detected if the reader's lesion annotations overlapped with the corresponding reference. Wilson 95% confidence intervals were computed for all sensitivity estimates. Statistical differences in sensitivity between Opt-40 and UD-40, and between Opt-60 and UD-60, were assessed using McNemar's paired test, with $p < 0.05$ considered statistically significant.

Aggregated sensitivity for malignant lesions and for all lesions was additionally evaluated as a function of lesion size and modeled using a binomial logistic regression. Bootstrap-based 95% confidence intervals were obtained by resampling lesions with replacement, refitting the model for each iteration, and extracting the 2.5th and 97.5th percentiles of the resulting distributions. From the fitted models, lesion size thresholds corresponding to 50% and 90% sensitivity were derived for each condition, along with their bootstrap confidence intervals.

Median Likert scores for abdominal image quality were calculated for each reader across the 22 patient cases. Differences in subjective image quality between UD-40 and Opt-40, and between UD-60 and Opt-60, were evaluated using the Wilcoxon signed-rank test, with $p < 0.05$ indicating statistical significance. All statistical analyses were performed in Python 3.10 using the SciPy 1.15 and Statsmodels 0.14 libraries. Data visualizations were generated in Python (Matplotlib 3.10.7) using code written with the assistance of generative AI coding tools (ChatGPT, GPT-6.2, OpenAI; Claude Opus 4.8, Anthropic).

This single-center, retrospective study was approved by our institutional review board and complied with the Health Insurance Portability and Accountability Act. The requirement for informed consent was waived.

## 3. Results

### *3.1. Patient Cohort*

The cohort consisted of 68 liver lesions (30 benign, 38 malignant) among 22 patients. The patients had the following primary sites of cancer: colorectal (n=7), lung (n=4), ovarian (n=2), stomach (n=3), breast (n=2), cervical (n=1), endometrial (n=1), lymphoma (n=1), and prostate (n=1). Benign liver lesions included simple cysts, hemangiomas, and calcified granulomas. Overall, patients had a median of 3 liver lesions (range: 0–9) with a median lesion size of 1.0 cm (range: 0.7–1.8 cm). By size, 31 lesions measured <1 cm, 21 lesions measured between 1-2 cm, and 16 lesions measured >2 cm. The patient cohort characteristics and their lesions are summarized in Table II.

### *3.2 Effective Dose*

Figure 2 and Supplementary Material S3 show the distribution of patient-level effective dose and organ-level doses contributing to the effective dose calculations across the different acquisition conditions. Median patient-level effective doses (Q1–Q3) were 7.8 (7.0–9.0) mSv, 12.0 (10.4–13.5) mSv, and 20.1 (17.4–22.5) mSv for Opt-40, Opt-60, and Orig-100, respectively. In comparison, median liver doses were 10.4 (9.3–12.4) mGy, 15.9 (13.4–19.9) mGy, and 18.8 (14.6–22.6) mGy, respectively, reflecting the preferential preservation of liver dose with task-specific organ-level TCM. Organ-level dose distributions for the remaining tissues contributing to the effective-dose calculation are provided in Supplementary Material S3.

### *3.3. Diagnostic Performance*

Detection rates for the three readers are summarized in Table III. Malignant lesions were detected at higher percentage rates with Opt-40 (mean = 68, range = 63–74) than with UD-40 (mean = 52, range = 47–55). Similarly, mean malignant lesion detection was higher with Opt-60 (mean = 82, range = 68–90) than UD-60 (mean = 67, range = 58–79), and was comparable to Orig-100 (mean = 80, range = 71–90). A similar trend was observed for detection of all lesions (malignant and benign combined). Mean detection rates increased from 54 (range = 47–60) with UD-40 to 64 (range = 59–69) with Opt-40, and from 65 (range = 57–72) with UD-60 to 77 (range = 69–84) with Opt-60. Mean all-lesion detection with Opt-60 was also comparable to Orig-100 (mean = 76, range = 72–82). McNemar's paired test showed statistically significant differences in sensitivity between UD-40 and Opt-40 and between UD-60 and Opt-60 for malignant lesions ($p = 0.0005$ and $p = 0.003$, respectively) and for all lesions (i.e. benign and malignant) ($p = 0.002$ and $p = 0.0002$, respectively).

### *3.4. Relationship of Diagnostic Performance with Lesion Size and Anatomical Location*

Figure 3 provides a visual depiction of the detection rates for the 68 liver lesions (38 malignant and 30 benign) along with their corresponding size, location, and benignity. Sensitivity values for the 8 liver segments and for the different acquisition methods are reported in Supplementary Material S4. Across liver segments, sensitivity followed the overall trend with Opt-40 higher than UD-40 and Opt-60 higher than UD-60, and similar between Opt-60 and Orig-100. No consistent segment-specific differences were observed within each dose condition.

Figure 4 shows lesion size versus sensitivity plots for malignant lesion detection across the five acquisition methods. Lesion-size thresholds corresponding to 50% and 90% sensitivity ($x_{50}$ and $x_{90}$) are summarized in Table IV. The fitted sensitivity curves yielded lower lesion-size thresholds for Opt-40 ($x_{50}$ = 8.7 mm; $x_{90}$ = 19.2 mm) than for UD-40 ($x_{50}$ = 13.6 mm; $x_{90}$ = 31.7 mm). Similarly, lower thresholds were observed for Opt-60 ($x_{50}$ = 3.3 mm; $x_{90}$ = 15.7 mm) than for UD-60 ($x_{50}$ = 7.9 mm; $x_{90}$ = 24.0 mm). Thresholds for Opt-60 were comparable to those estimated for Orig-100 ($x_{50}$ = 3.7 mm; $x_{90}$ = 16.8 mm).

A similar trend was observed for the combined sensitivity of all liver lesions (i.e., both benign and malignant lesions). As shown on Table IV, smaller lesions were detectable with higher sensitivity in Opt-40 ($x_{50}$ = 8.1 mm, $x_{90}$ = 17.8 mm) than in its uniform dose reduction counterpart, UD-40 ($x_{50}$ = 11.1 mm, $x_{90}$ = 25.4 mm). This can also be observed for Opt-60 ($x_{50}$ = 4.2 mm, $x_{90}$ = 13.7 mm) compared with UD-60 ($x_{50}$ = 7.5 mm, $x_{90}$ = 18.4 mm). Similarly, lesion size thresholds corresponding to 50% and 90% sensitivity were comparable between Opt-60 ($x_{50}$ = 4.2 mm, $x_{90}$ = 13.7 mm) and Orig-100 ($x_{50}$ = 3.6 mm, $x_{90}$ = 15.7 mm). Supplementary Material S5 shows lesion size versus sensitivity plots for detection of all lesions (benign and malignant) across all five acquisition methods.

### *3.5. Image Quality*

The median abdominal image quality ratings for UD-40 across the three readers ranged from suboptimal with compromised diagnostic confidence, (Likert Score = 2) to acceptable diagnostic quality (Likert Score = 3). In contrast, median ratings for Opt-40 ranged from diagnostic with acceptable image quality (Likert Score = 3) to good image quality (Likert Score = 4). UD-60 images were consistently rated by all readers as diagnostic with acceptable image quality (Likert score = 3). Opt-60 images received higher quality ratings, ranging from acceptable to good diagnostic image quality (Likert score = 3–4.5). Orig-100 received the highest median abdominal image quality ratings, ranging from good to excellent (Likert Score = 4-5). Figure 5 summarizes the abdominal image quality ratings assigned by the readers across the different acquisition methods. Based on the Wilcoxon Signed-Rank Test, differences in median Likert Score ratings between UD-40 and Opt-40 were statistically significant for Readers 1 and 3 ($p \leq 0.012$) but not for Reader 2 ($p = 0.24$). Differences between UD-60 and Opt-60 were statistically significant for all readers ($p \leq 0.0067$).

Figures 6 and 7 show representative cases illustrating the differences in abdominal image quality and lesion detection from uniform (UD) to optimized (Opt) dose reduction across the UD-40, Opt-40, UD-60, Opt-60, and Orig-100 series. In both figures, image quality improved in that order, as did lesion detection.

Relative to uniform dose reduction, pelvic image quality was comparable at the 40% dose level and lower at the 60% level. Median pelvic ratings for UD-40 and Opt-40 ranged from suboptimal with compromised diagnostic confidence to acceptable diagnostic quality (Likert = 2–3). At the 60% dose level, UD-60 ratings ranged from suboptimal–acceptable to good (Likert = 2.5–4), whereas Opt-60 ranged from suboptimal to acceptable (Likert = 2–3). Orig-100 received the highest pelvic ratings (Likert = 4–5). Supplementary Material S6 summarizes the pelvic image quality ratings assigned by the readers across the different acquisition methods. Based on the Wilcoxon signed-rank test, pelvic image quality did not differ significantly between UD-40 and Opt-40 for Readers 1 and 3 ($p = 0.83$ and $0.37$) but was significantly lower with Opt-40 for Reader 2 ($p = 0.03$). Between UD-60 and Opt-60, it was significantly lower with Opt-60 for Readers 1 and 2 ($p = 0.02$ and $0.03$) but not for Reader 3 ($p = 0.10$).

## 4. Discussion

In this study, we demonstrate that retroOpt, a retrospective organ-level tube current modulation (TCM) algorithm, improved hepatic lesion detection compared with uniform dose reduction at equivalent patient-specific effective-dose targets. Unlike conventional TCM methods that primarily adjust tube current according to patient attenuation, retroOpt performs task-specific optimization to preferentially preserve image quality within organs of interest while maintaining a predefined effective dose constraint.[18] At 60% of the original effective dose, mean malignant lesion detection sensitivity among our three readers was higher with Opt-60 than with UD-60 (82% vs 67%) and was comparable to that of the original clinical acquisition (80%). This pattern was also observed in our subjective abdominal image quality results, where optimized acquisitions received higher ratings than their uniform dose-reduction counterparts. Furthermore, the fitted lesion-size thresholds corresponding to 50% and 90% sensitivity were lower with optimized than with uniform dose reduction at the two evaluated dose levels. For Opt-60, image quality and detection thresholds were comparable to those observed at full dose.

This ability to reduce radiation dose while preferentially preserving hepatic lesion detectability is especially important in liver imaging. The liver is among the most common organs of metastatic spread, and metastatic disease accounts for about 90% of cancer-related deaths.[1–3,25] Consequently, repeated contrast-enhanced CTs are routinely performed for detection and surveillance of hepatic metastases. However, because liver lesions frequently demonstrate only subtle attenuation differences relative to surrounding parenchyma, detection is particularly vulnerable to increased image noise at reduced dose levels.

Prior studies have attempted to define dose levels at which reliable liver lesion detection remains feasible. Studies using filtered back projection reconstruction have demonstrated that liver metastasis detection is largely preserved at moderate dose reductions of 25–50% ($CTDI_{vol}$: 8.1–10.4 mGy) but deteriorates substantially at more aggressive reductions corresponding to $CTDI_{vol}$ levels of 1.7–5.0 mGy.[6,22,23,26–28] Our results with uniform dose reduction are consistent with this trend. Relative to Orig-100, sensitivity loss at UD-60 ($CTDI_{vol}$ = 7.7 mGy) was less pronounced than at UD-40 ($CTDI_{vol}$ = 5.2 mGy). This behavior is consistent with findings from Fält et al., who reported progressively larger sensitivity losses (10%-20%) at $CTDI_{vol}$ levels below approximately 8.6 mGy.[29]

RetroOpt enables liver lesion detection to be maintained at lower effective-dose levels compared to conventional strategies. Mean malignant lesion detection sensitivity for Opt-40 (68%) was similar to that of UD-60 (67%), despite corresponding to 40% rather than 60% of the original effective dose. Similarly, sensitivity for Opt-60 (82%) was similar to that of Orig-100 (80%), despite a 40% reduction in effective dose. Task-specific organ-level optimization may enable greater effective-dose reductions than uniform dose reduction while preferentially preserving hepatic lesion detectability.

Smaller lesions are particularly susceptible to image quality degradation. Fletcher et al. previously reported reduced reader confidence for malignant hepatic lesions smaller than 15 mm.[10,22] In our study, the estimated lesion-size thresholds corresponding to 50% and 90% sensitivity were similar between Opt-60 ($x_{50}$ = 3.3 mm, $x_{90}$ = 15.7 mm) and Orig-100 ($x_{50}$ = 3.7 mm, $x_{90}$ = 16.8 mm). Logistic regression models also found higher sensitivity for lesions in the 5–15 mm range with Opt-60 compared with that of other reduced-dose strategies. These findings suggest that task-specific organ-level optimization preserve detection of smaller metastases. Furthermore, consistent with prior work,[29] we did not identify a consistent relationship between lesion location and detection sensitivity across dose-reduction strategies.

Previous studies of organ-level TCM found that noise levels and subjective image quality within the liver and adjacent organs could be preserved at moderately reduced overall radiation dose.[14,15] However, these studies did not evaluate whether these improvements resulted in improved diagnostic task performance. We sought to address this gap and found that retroOpt achieved acceptable-to-good abdominal image quality at Opt-60 (Likert scale 3–4.5) and improved hepatic lesion detection compared with uniform dose reduction at the same effective-dose target. By preferentially preserving radiation dose within the liver, retroOpt may improve the image noise characteristics relevant to lesion-background discrimination, an important determinant of detectability for low-contrast hepatic lesions.[30–32]

Besides evaluating diagnostic task performance, our study also differs from prior work by incorporating patient-specific organ dosimetry into the optimization framework. While previous organ-level TCM approaches have primarily used scanner-based dose metrics such as $CTDI_{vol}$, retroOpt uses an ICRP 103-based weighted combination of patient-specific organ doses to define the effective-dose constraint.[13] Patient-specific organ doses were estimated using Monte Carlo simulations and anatomical segmentations, allowing the optimization to account for differences in patient anatomy and organ dose distribution. Within this predefined effective-dose constraint, retroOpt minimizes weighted organ-level image noise, with preferential weighting assigned to the liver to preserve image quality for hepatic lesion detection. This redistribution was reflected in our dosimetric results: at Opt-60, median effective dose decreased from 20.1 to 12.1 mSv relative to Orig-100, while median liver dose decreased more modestly from 18.8 to 15.9 mGy. These findings demonstrate how task-specific organ-level TCM can preferentially preserve radiation dose within the organ of interest while reducing the overall effective-dose metric.

Because retroOpt allows the organ-level noise weighting $w_L^n$ to be tuned to the diagnostic task, it supports a range of configurations that trade overall dose reduction against image quality in non-target regions. The configuration evaluated here prioritized the liver, with weights set so that liver dose remained within ranges previously associated with adequate hepatic lesion detection[6,10,22,23] while effective dose was reduced to 40% and 60% of the original. This reallocation improved hepatic detection, while pelvic image quality was comparable to or lower than that of uniform dose reduction. Prior liver-targeted organ-based modulation preserved image quality across thoraco-abdominopelvic regions, including the pelvis, but at more modest reductions (~23–31% CTDIvol).[15] This represents a more conservative point on the same tradeoff. The configuration studied here is best suited to liver-focused indications. Extending retroOpt to tasks in which the pelvis is also a diagnostic target would require a weighting matched to that combined task. Whether such a weighting can maintain a comparable overall dose reduction depends on the relative dose requirements of the diagnostic tasks involved. Characterizing this tradeoff, by sweeping weighting configurations and extending task-based assessment to extrahepatic and pelvic pathology, is an important direction for future work.

Our study had several limitations. First, it was a single-center, retrospective study performed on a single CT scanner platform. Generalizability will require validation in larger, multi-center cohorts. Second, the cohort was relatively small (22 patients, 68 lesions), and only three radiologists participated in the reader study. Furthermore, the statistical analyses did not explicitly account for clustering of lesions within patients or repeated evaluations across readers. The primary goal, however, was to evaluate whether organ-level dose optimization through retroOpt improved hepatic lesion detection relative to uniform dose reduction at equivalent effective-dose targets, rather than to establish definitive clinical performance estimates. Third, dose reduction was simulated using projection-domain noise insertion rather than prospectively acquired low-dose images, enabling comparison across dose levels without additional patient exposure. Although the noise modeling framework and optimization approach have undergone previous phantom-based validation,[18,19] prospective clinical validation is necessary. Fourth, evaluation was limited to filtered back projection reconstruction. Advanced iterative and deep learning–based reconstruction methods have demonstrated promise at reduced dose,[10,11] and their combination with retroOpt warrants further investigation. Fifth, because this implementation of retroOpt was optimized for hepatic lesion detection, dose was redistributed away from non-target regions, and pelvic image quality was maintained or reduced relative to uniform dose reduction.

Diagnostic-task performance for pathology outside the liver, including in the pelvis, was not evaluated at these reduced doses. Weighting schemes matched to such tasks are a direction for future work. Finally, the current retroOpt implementation is retrospective. Prospective organ-level modulation would require real-time organ localization and dose estimation prior to the CT scan, e.g., from scout images. Recent advances in deep learning suggest this is feasible,[33–35] and its prospective implementation is currently under development within our group as an important step toward clinical translation.

In summary, the proposed organ-level TCM approach improved hepatic lesion detection compared with uniform dose reduction at equivalent effective-dose targets, while preferentially preserving image quality within the liver. These findings suggest that task-specific organ-level optimization may enable greater effective-dose reductions than conventional uniform dose reduction strategies while maintaining hepatic lesion detectability.

## 5. Declaration of generative AI and AI-assisted technologies in the manuscript preparation process

During the preparation of this work, the authors used ChatGPT (GPT-6.2, OpenAI) and Claude Opus 4.8 (Anthropic) to assist with language editing and with revising the clarity, structure, and organization of the manuscript text. After using these tools, the authors reviewed and edited the content as needed and take full responsibility for the content of the published article.

## 7. Tables

**Table I:** Acquisition and reconstruction parameters.

| Acquisition Parameters | Values |
|---|---|
| Tube voltage (number of patients) | |
| 100 kVp | 5 (22.7%) |
| 120 kVp | 16 (72.7%) |
| 140 kVp | 1 (4.5%) |
| Pitch (number of patients) | |
| 0.99 | 21 (95.5%) |
| 0.51 | 1 (4.5%) |
| Median Rotation Time (IQR) | 0.65 (0.50–0.80) s |
| Median Noise Index (IQR) | 12.4 (11.5–13.0) |
| Median $CTDI_{vol}$ (IQR) | 12.9 (10.1–16.3) mGy |
| Field-of-View | 500 mm |
| Collimation | 80 mm |
| Section Thickness | 2.5 mm |
| Matrix Size | 512x512 pixels |
| Reconstruction Kernel | Standard |
| Reconstruction | Filtered Backprojection |

**Table II:** Patient baseline characteristics. Values report absolute number with interquartile range or percentages in parentheses.

| Parameters | Values |
|---|---|
| **Patients** | |
| Number of patients | 22 |
| Median Age (IQR) | 57 (43–65) years |
| **Sex** | |
| Male | 10 (45%) |
| Female | 12 (55%) |
| Median BMI (IQR) | 25.8 (22.0-29.3) kg/$m^2$ |
| **Cancer Types** | |
| -Colorectal | 7 (31.8%) |
| -Lung | 4 (18.2 %) |
| -Stomach | 3 (13.6%) |
| -Breast | 2 (9.1%) |
| -Ovarian | 2 (9.1%) |
| -Cervical | 1 (4.5%) |
| -Endometrial | 1 (4.5%) |
| -Lymphoma | 1 (4.5%) |
| -Prostate | 1 (4.5%) |
| **Lesions** | |
| Number of lesions | 68 |
| Malignant | 38 (55.9%) |
| Benign | 30 (44.1%) |
| **Number of lesions per patient** | |
| 0 | 2 (9.1%) |
| 1-2 | 7 (31.8 %) |
| 3-5 | 9 (40.9%) |
| >5 | 4 (18.2%) |
| Median Lesion Size (IQR) | 1.0 (0.7–1.8) cm |
| **Lesion size distribution** | |
| <1 cm | 31 (45.6%) |
| 1-2 cm | 21 (30.9%) |
| >2 cm | 16 (23.5%) |

**Table III:** Sensitivity performance for different dose reduction methods and reduction levels.

| Mean Sensitivity for Malignant Lesions [Confidence Intervals] (%) | | | | | |
|---|---|---|---|---|---|
| | **UD-40** | **Opt-40** | **UD-60** | **Opt-60** | **Orig-100** |
| **R1** | 47 [31 –64] | 63 [46 –78] | 58 [41 –74] | 68 [51 –83] | 71 [54 –85] |
| **R2** | 55 [38 –71] | 66 [49 –80] | 63 [46 –78] | 87 [72 –96] | 79 [63 –90] |
| **R3** | 53 [36 –69] | 74 [57 –87] | 79 [63 –90] | 90 [75 –97] | 90 [75 –97] |
| **Mean Sensitivity for All Lesions [Confidence Intervals] (%)** | | | | | |
| | **UD-40** | **Opt-40** | **UD-60** | **Opt-60** | **Orig-100** |
| **R1** | 47 [35 –60] | 59 [46 –71] | 57 [45 –69] | 69 [57 –80] | 72 [60 –82] |
| **R2** | 60 [48 –72] | 63 [51 –75] | 66 [54 –77] | 79 [68 –88] | 75 [63 –85] |
| **R3** | 55 [42 –67] | 69 [57 –80] | 72 [60 –82] | 84 [73 –92] | 82 [71 –91] |

**Table IV:** Lesion-size thresholds and corresponding confidence intervals for malignant and all lesions.

| Sensitivity Size for Malignant Lesions (mm) [Confidence Intervals] | | | | | |
|---|---|---|---|---|---|
| | **UD-40** | **Opt-40** | **UD-60** | **Opt-60** | **Orig-100** |
| **50%** | 13.6 [7.22, 17.8] | 8.65 [4.68, 11.2] | 7.91 [-0.86, 11.4] | 3.33 [-8.0, 7.12] | 3.69 [-15.8, 7.84] |
| **90%** | 31.7 [23.0, 44.3] | 19.2 [13.4, 28.0] | 24.0 [15.2, 44.5] | 15.7 [10.4, 23.4] | 16.8 [9.95, 29.0] |
| **Sensitivity Size for All Lesions (mm) [Confidence Intervals]** | | | | | |
| | **UD-40** | **Opt-40** | **UD-60** | **Opt-60** | **Orig-100** |
| **50%** | 11.1 [7.95, 14.0] | 8.05 [5.85, 9.83] | 7.53 [4.59, 9.54] | 4.17 [-0.61, 6.07] | 3.62 [-3.78, 6.32] |
| **90%** | 25.4 [18.3, 32.1] | 17.8 [13.0, 24.3] | 18.4 [13.2, 26.9] | 13.7 [9.90, 20.2] | 15.7 [10.4, 23.8] |

## 8. Figures Captions

**FIGURE 1:** Flowchart of the patient selection process, with inclusion and exclusion criteria.

**FIGURE 2:** Box-and-whisker plots showing the distribution of organ-level dose (mGy) and overall effective dose (mSv) for Orig-100, Opt-40, and Opt-60 across the 22 patients in our study. Dose results are presented from highest to lowest mean absolute dose.

**FIGURE 3**: Schematic of the anatomic locations of all malignant and benign liver lesions across all five acquisition methods, with corresponding lesion type and size. Grayscale shading denotes the number of readers who detected each lesion. Lesions were mapped to their corresponding location defined as the caudate lobe (segment I), left lateral sector (segments II-III), left medial sector (segment IV), right anterior sector (segments V and VIII), and right posterior sector (segments VI and VII). Lesions annotated as adjacent to the capsule or fissure were within 1 cm of either structure.

**FIGURE 4**: Lesion size versus sensitivity plots for malignant lesion detection across all five acquisition methods. Blue dots denote sensitivity per lesion, the blue line corresponds to the binomial logistic regression fit, and the orange overlay represents the 95% confidence interval. Vertical dotted lines indicate the lesion sizes $x_{50}$ and $x_{90}$, corresponding to malignant lesion sensitivities of 50% and 90%, respectively. Negative lower bounds on the confidence intervals are a mathematical artifact of the logistic regression.

**FIGURE 5**: Violin plot summarizing the readers' image quality scores for each acquisition method for the abdomen. The white-filled circle denotes the median score, and the width and number of dots represent the number of responses for each image quality score.

**FIGURE 6**: Representative images of a malignant liver lesion (solid white arrow), emulated with (A) UD-40, (B) Opt-40, (C) UD-60, (D) Opt-60, and (E) Orig-100. The liver lesion was detected by no readers on UD-40 or Opt-40, by 2 readers on UD-60, by all 3 readers on Opt-60, and by 2 readers on Orig-100. Median Likert image quality scores of 3, 4, 3, 4, and 4 were assigned to panels (A)–(E), respectively.

**FIGURE 7**: Representative images of two differently sized malignant liver lesions (2.8-cm lesion, solid black arrow; 1.2-cm lesion, solid white arrow), emulated with (A) UD-40, (B) Opt-40, (C) UD-60, (D) Opt-60, and (E) Orig-

100. A zoomed-in view of the 1.2-cm lesion is shown in the bottom-right inset (yellow outline). The 2.8-cm lesion was detected by 2 readers on UD-40 and all 3 readers on Opt-40, UD-60, Opt-60, and Orig-100 images. The 1.2-cm lesion was detected by no readers on UD-40, by 2 readers on Opt-40, by all three readers on UD-60 and Opt-60, and by 2 readers on Orig-100. Median Likert image quality scores of 2, 3, 3, 4, and 5 were assigned to panels (A)–(E), respectively.

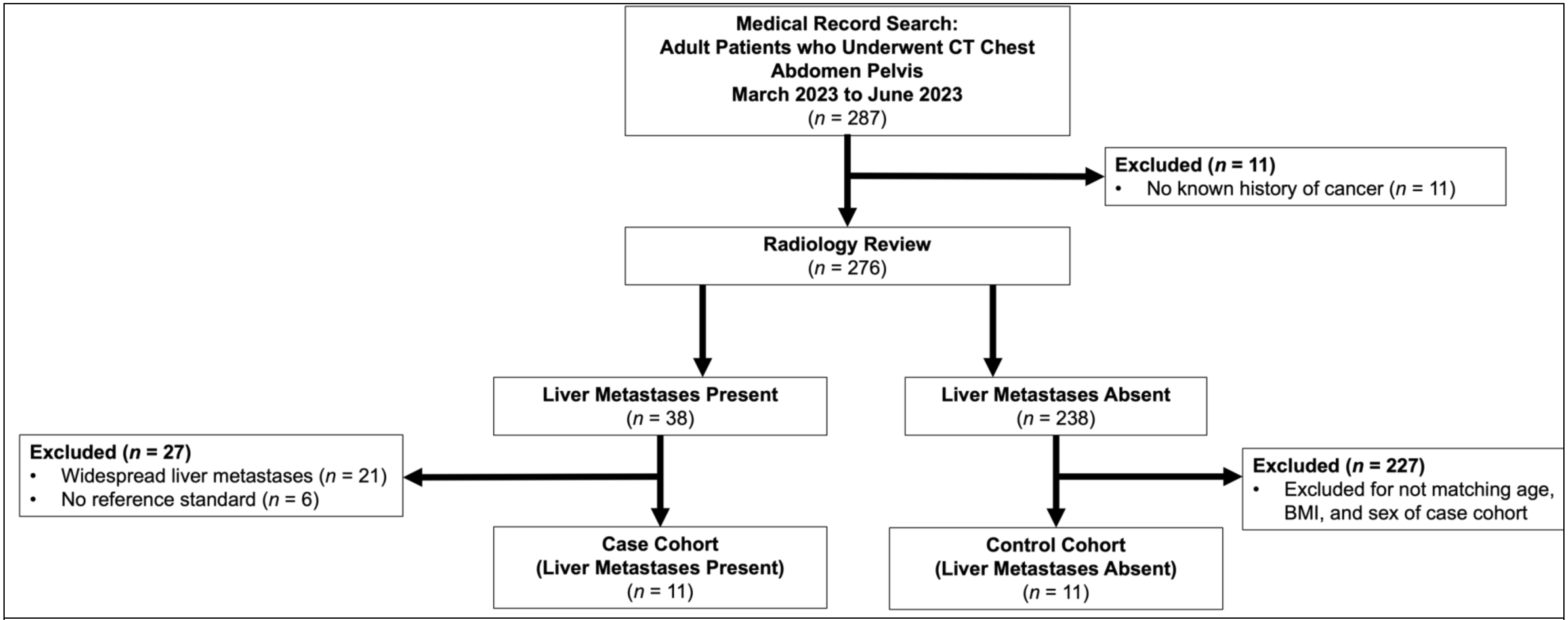

**FIGURE 1**: Flowchart of the patient selection process, with inclusion and exclusion criteria.

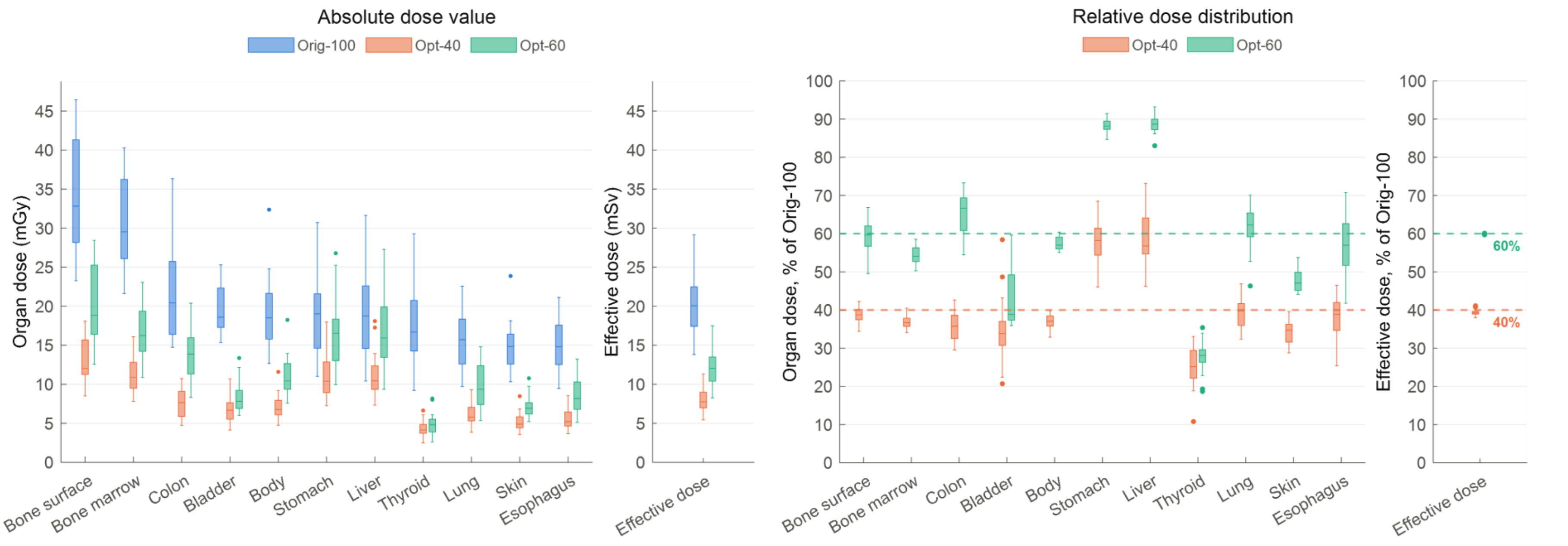

**FIGURE 2**: Box-and-whisker plots showing the distribution of organ-level dose (mGy) and overall effective dose (mSv) for Orig-100, Opt-40, and Opt-60 across the 22 patients in our study. Dose results are presented from highest to lowest mean absolute dose.

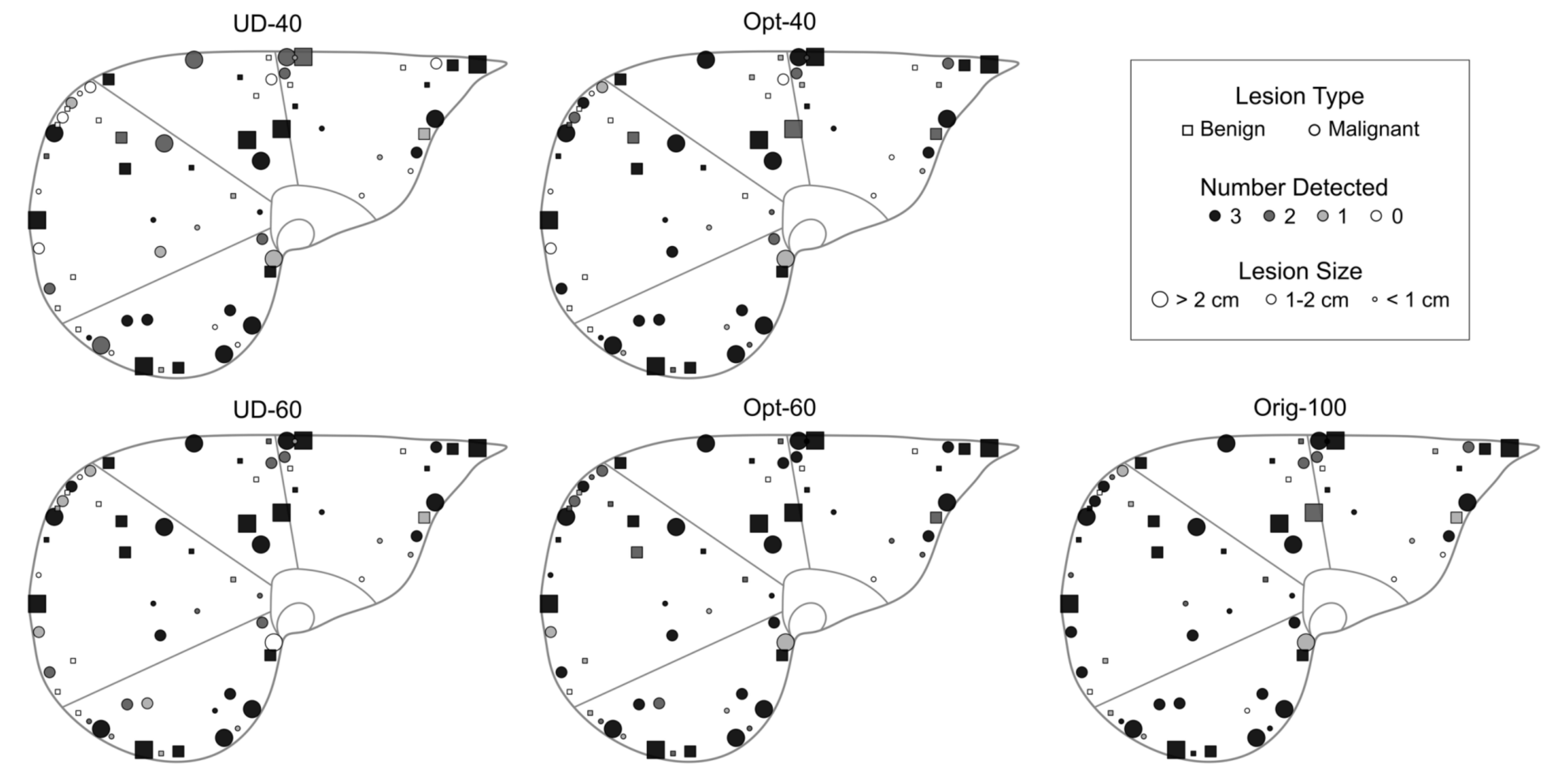

**FIGURE 3**: Schematic of the anatomic locations of all malignant and benign liver lesions across all five acquisition methods, with corresponding lesion type and size. Grayscale shading denotes the number of readers who detected each lesion. Lesions were mapped to their corresponding location defined as the caudate lobe (segment I), left lateral sector (segments II-III), left medial sector (segment IV), right anterior sector (segments V and VIII), and right posterior sector (segments VI and VII). Lesions annotated as adjacent to the capsule or fissure were within 1 cm of either structure.

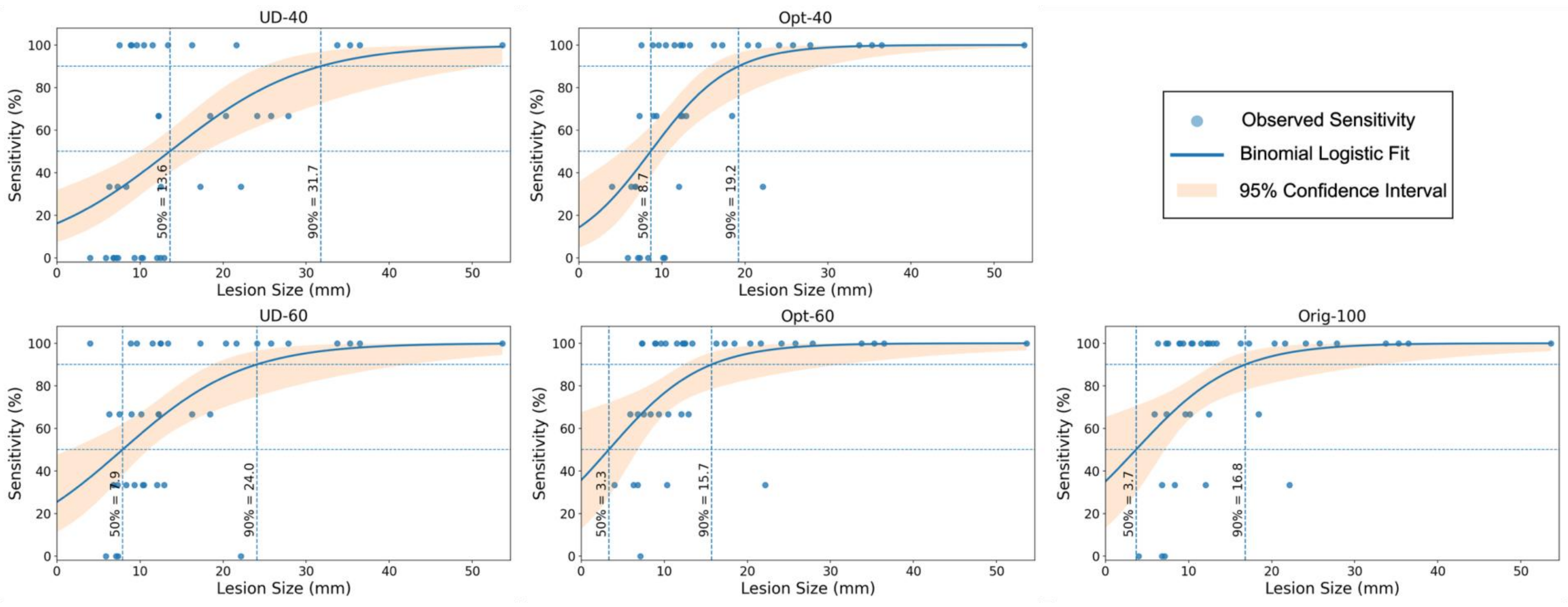

**FIGURE 4**: Lesion size versus detection sensitivity plots for malignant lesion detection across all five acquisition methods. Blue dots denote sensitivity per lesion, the blue line corresponds to the binomial logistic regression fit, and the orange overlay represents the 95% confidence interval. Vertical dotted lines indicate the lesion sizes $x_{50}$ and $x_{90}$, corresponding to malignant lesion sensitivities of 50% and 90%, respectively. Negative lower bounds on the confidence intervals are a mathematical artifact of the logistic regression.

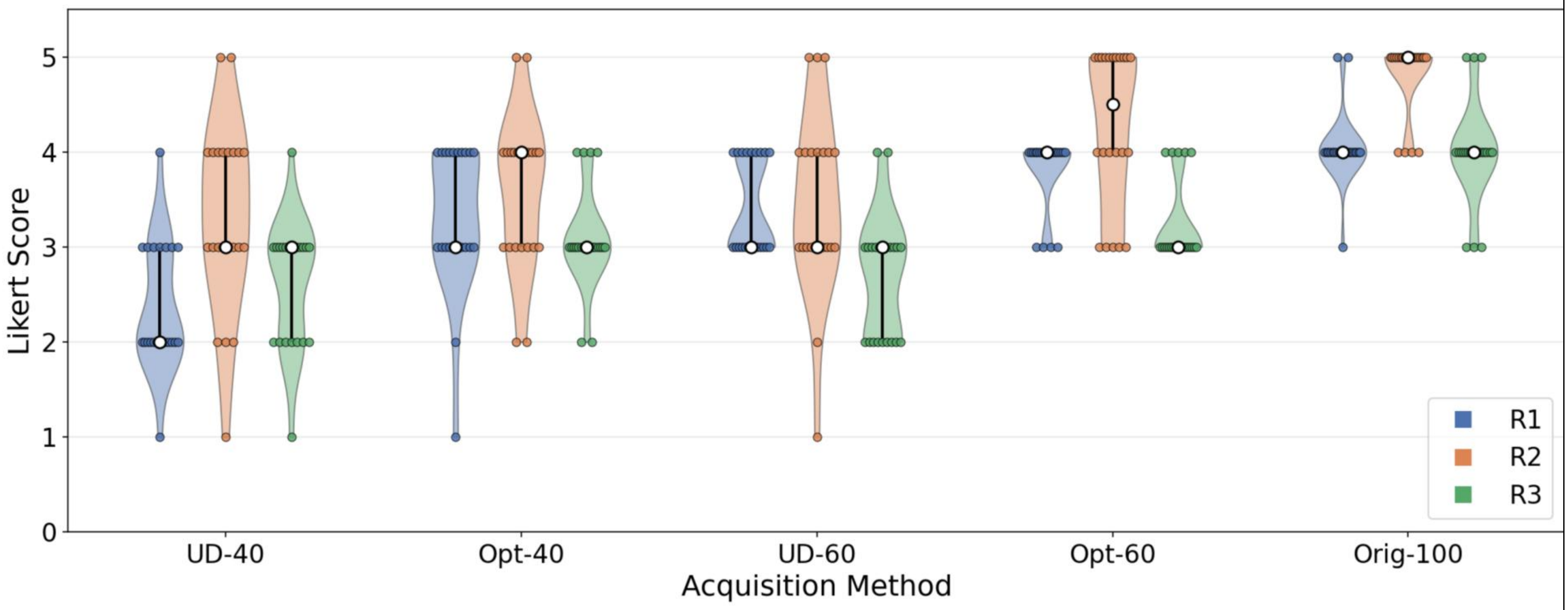

**FIGURE 5**: Violin plot summarizing the readers' image quality scores for each acquisition method for the abdomen. The white-filled circle denotes the median score, and the width and number of dots represent the number of responses for each image quality score.

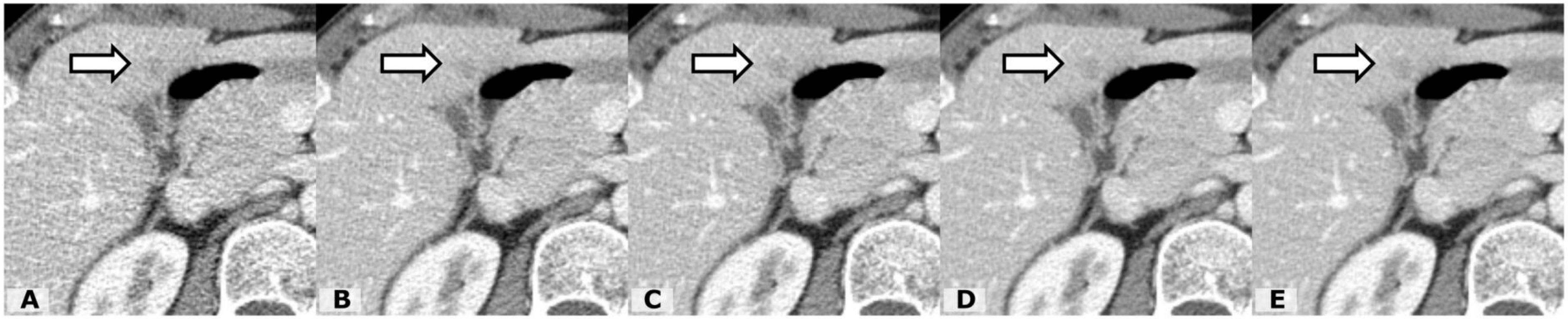

**FIGURE 6**: Representative images of a malignant liver lesion (solid white arrow), emulated with (A) UD-40, (B) Opt-40, (C) UD-60, (D) Opt-60, and (E) Orig-100. The liver lesion was detected by no readers on UD-40 or Opt-40, by 2 readers on UD-60, by all 3 readers on Opt-60, and by 2 readers on Orig-100. Median Likert image quality scores of 3, 4, 3, 4, and 4 were assigned to panels (A)–(E), respectively.

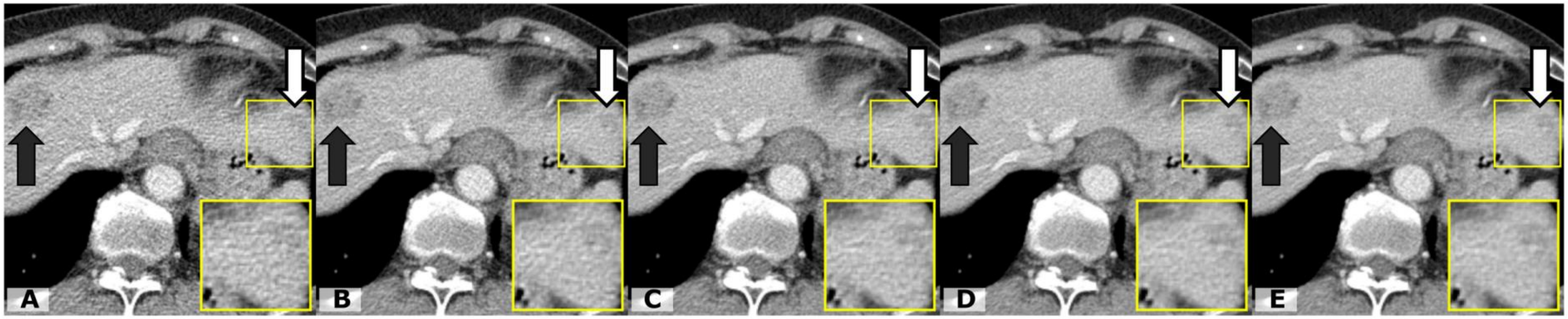

**FIGURE 7**: Representative images of two differently sized malignant liver lesions (2.8-cm lesion, solid black arrow; 1.2-cm lesion, solid white arrow), emulated with (A) UD-40, (B) Opt-40, (C) UD-60, (D) Opt-60, and (E) Orig-100. A zoomed-in view of the 1.2-cm lesion is shown in the bottom-right inset (yellow outline). The 2.8-cm lesion was detected by 2 readers on UD-40 and all 3 readers on Opt-40, UD-60, Opt-60, and Orig-100 images. The 1.2-cm lesion was detected by no readers on UD-40, by 2 readers on Opt-40, by all three readers on UD-60 and Opt-60, and by 2 readers on Orig-100. Median Likert image quality scores of 2, 3, 3, 4, and 5 were assigned to panels (A)–(E), respectively.

# Supplementary Material

## S1. Retrospective Organ-Level Tube Current Modulation

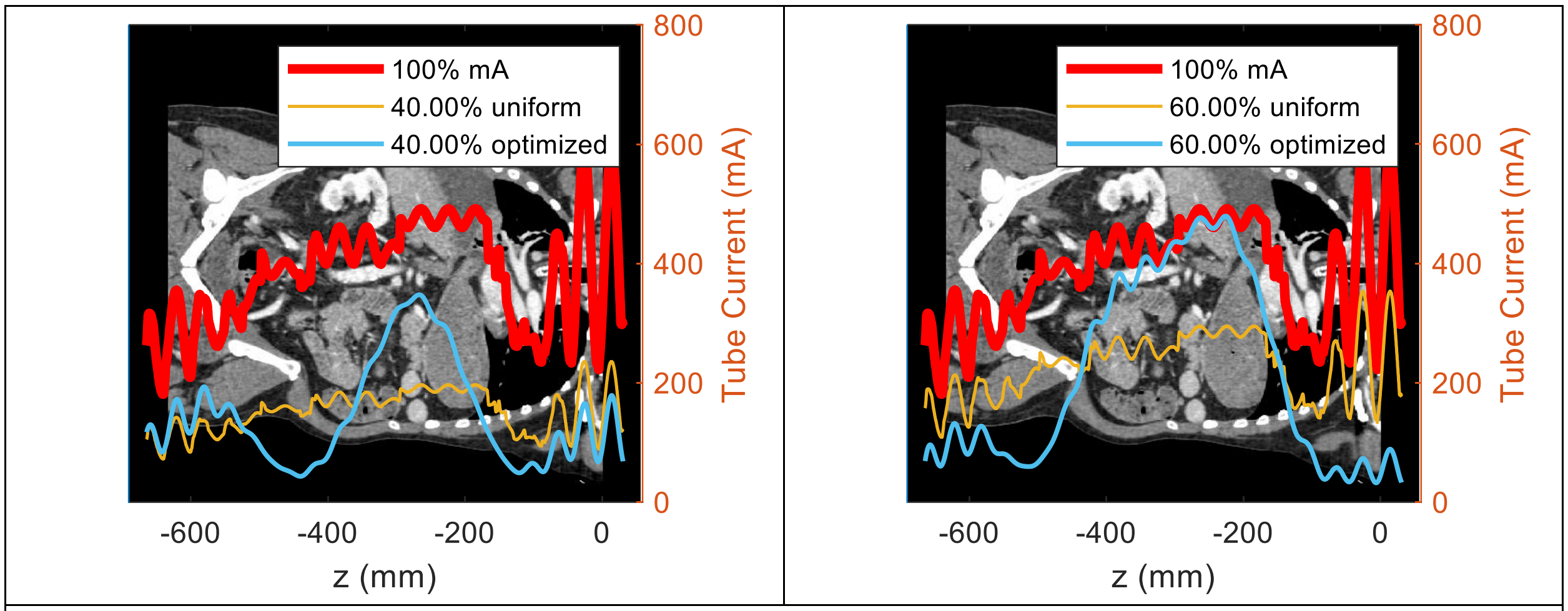

**Figure S1:** Sample retrospectively optimized tube current modulation profiles designed for optimal image quality at the liver

## S2. Likert Score Details

**Table S1:** Detailed description of Likert Score ratings.

| | |
|---|---|
| **5** | **Diagnostic – Excellent Image Quality**<br>- Excellent overall image quality<br>- Very low image noise and reader highly preferred image texture<br>- Very high resolution to depict details with very high confidence.<br>- Very good contrast to differentiate between materials with very high confidence.<br>- Clean images with almost zero image artifacts |
| **4** | **Diagnostic – Good Image Quality**<br>- Good overall image quality<br>- Low image noise and reader preferred image texture<br>- High resolution to depict details with confidence.<br>- Good contrast to differentiate between materials with confidence.<br>- Minor artifacts not interfering with diagnostic decision-making. |
| **3** | **Diagnostic – Acceptable Image Quality**<br>- Acceptable overall image quality<br>- Moderate image noise and acceptable image texture<br>- Acceptable resolution to depict details with acceptable level of confidence.<br>- Acceptable contrast to differentiate between materials with an acceptable level of confidence<br>- Moderate artifacts possibly interfering with diagnostic decision-making. |
| **2** | **Sub-optimal diagnostic with compromised diagnosis confidence**<br>- Poor overall image quality<br>- Severe image noise and poor image texture that can affect diagnosis.<br>- Poor resolution to compromise confidence in depicting details.<br>- Poor contrast to differentiate between materials. |
| | - Major artifacts affecting visualization of major structures (diagnosis still possible) |
| **1** | **Non-diagnostic**<br>- Unacceptable overall image quality<br>- Unacceptably severe image noise and unacceptable image texture that affects diagnosis.<br>- Unacceptably poor resolution<br>- Unacceptable contrast to differentiate between materials.<br>- Unacceptable artifacts clearly affecting diagnosis. (non-diagnosis) |

## S3. Distribution of Organ-Level and Effective Doses

**Table S2:** Median organ-level doses with corresponding interquartile range for Orig-100, Opt-40, and Opt-60

| | **Orig-100**<br>**Median [Q1-Q3]** | **Opt-40**<br>**Median [Q1-Q3]** | **Opt-60**<br>**Median [Q1-Q3]** |
|---|---|---|---|
| **Bone surface** | 32.82 [28.17 – 41.32] | 12.02 [11.26 – 15.68] | 18.83 [16.38 – 25.26] |
| **Bone marrow** | 29.53 [26.09 – 36.24] | 10.88 [9.50 – 12.81] | 16.24 [14.22 – 19.37] |
| **Colon** | 20.42 [16.38 – 25.74] | 7.65 [5.89 – 9.10] | 13.86 [11.32 – 15.97] |
| **Bladder** | 18.60 [17.29 – 22.32] | 6.68 [5.52 – 7.65] | 7.82 [6.90 – 9.23] |
| **Body** | 18.51 [15.79 – 21.64] | 6.77 [6.09 – 7.94] | 10.43 [9.35 – 12.65] |
| **Stomach** | 19.02 [14.62 – 21.59] | 10.39 [8.94 – 12.86] | 16.53 [13.03 – 18.35] |
| **Liver** | 18.76 [14.58 – 22.60] | 10.44 [9.35 – 12.36] | 15.94 [13.44 – 19.91] |
| **Thyroid** | 16.68 [14.26 – 20.74] | 4.15 [3.73 – 4.85] | 4.81 [3.90 – 5.52] |
| **Lung** | 15.72 [12.59 – 18.34] | 5.79 [5.31 – 7.05] | 9.35 [7.41 – 12.38] |
| **Skin** | 14.83 [12.58 – 16.41] | 4.89 [4.41 – 5.84] | 6.93 [6.20 – 7.64] |
| **Esophagus** | 14.80 [12.52 – 17.58] | 5.24 [4.66 – 6.45] | 8.17 [6.78 – 10.29] |
| **Effective Dose** | 20.07 [17.42 – 22.48] | 7.75 [6.98 – 9.00] | 12.05 [10.40 – 13.49] |

## S4. Sensitivity for Different Lesion Locations

**Table S3:** Detection sensitivity of benign and malignant lesions across liver segments. No lesions were in the caudate lobe (segment I); detection across segments II–VIII is grouped into anatomic sectors.

| **Mean Sensitivity for All Lesions Across Different Liver Segments (%)** | | | | | |
|---|---|---|---|---|---|
| **Segment** | **UD-40** | **Opt-40** | **UD-60** | **Opt-60** | **Orig-100** |
| **II/III** | 56 [41–69] | 65 [51–77] | 67 [53–79] | 78 [64–88] | 69 [54–81] |
| **IV a/b** | 63 [42–81] | 59 [39–78] | 82 [62–94] | 85 [66–96] | 78 [58–91] |
| **V/VIII** | 42 [30–54] | 56 [43–67] | 57 [45–69] | 75 [63–85] | 78 [66–87] |
| **VI/VII** | 65 [50–78] | 77 [63–87] | 67 [52–79] | 77 [63–87] | 82 [69–92] |

## S5. Sensitivity vs. Lesions Size – All Lesions

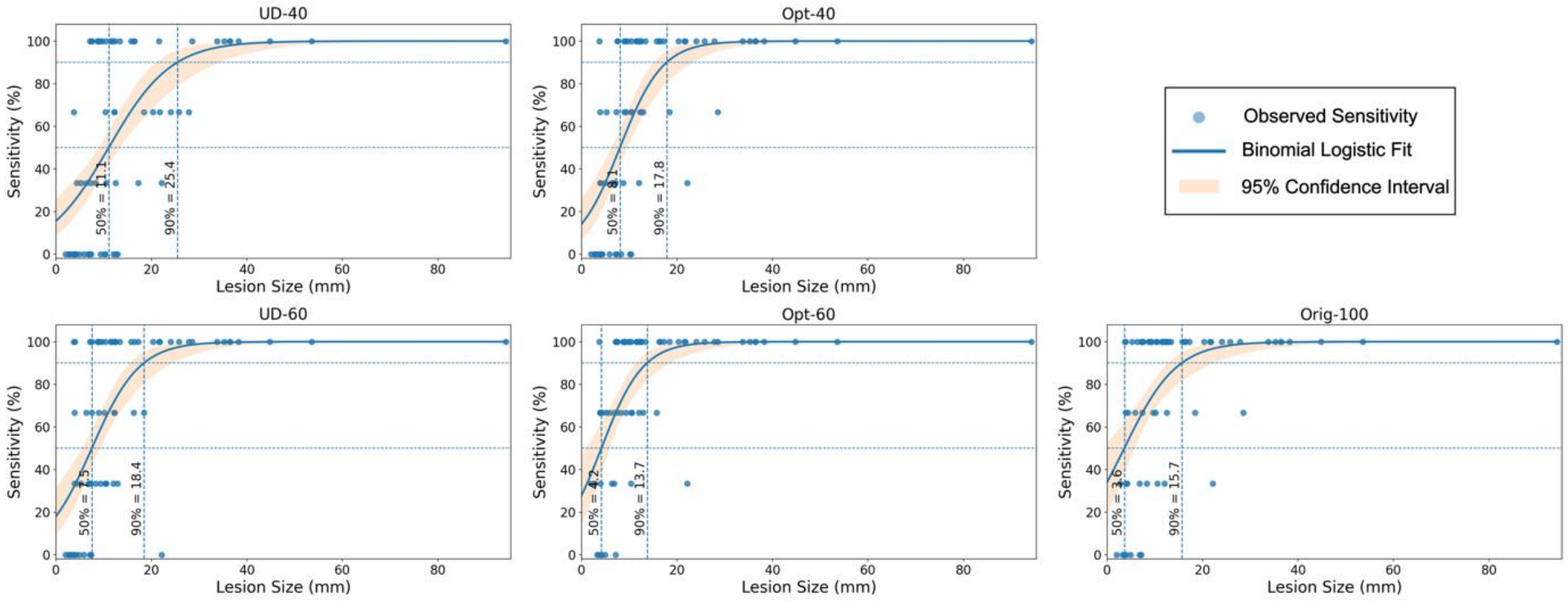

**Figure S2:** Lesion size versus sensitivity plots for all lesion detection (malignant and benign) across all five acquisition methods. Blue dots denote sensitivity per lesion, the blue line corresponds to the binomial logistic regression fit, and the orange overlay represents the 95% confidence interval. Vertical dotted lines indicate the lesion sizes $x_{50}$ and $x_{90}$, corresponding to all-lesion sensitivities of 50% and 90%, respectively. Negative lower bounds on the confidence intervals are a mathematical artifact of the logistic regression.

## S6. Image Quality Ratings – Pelvis

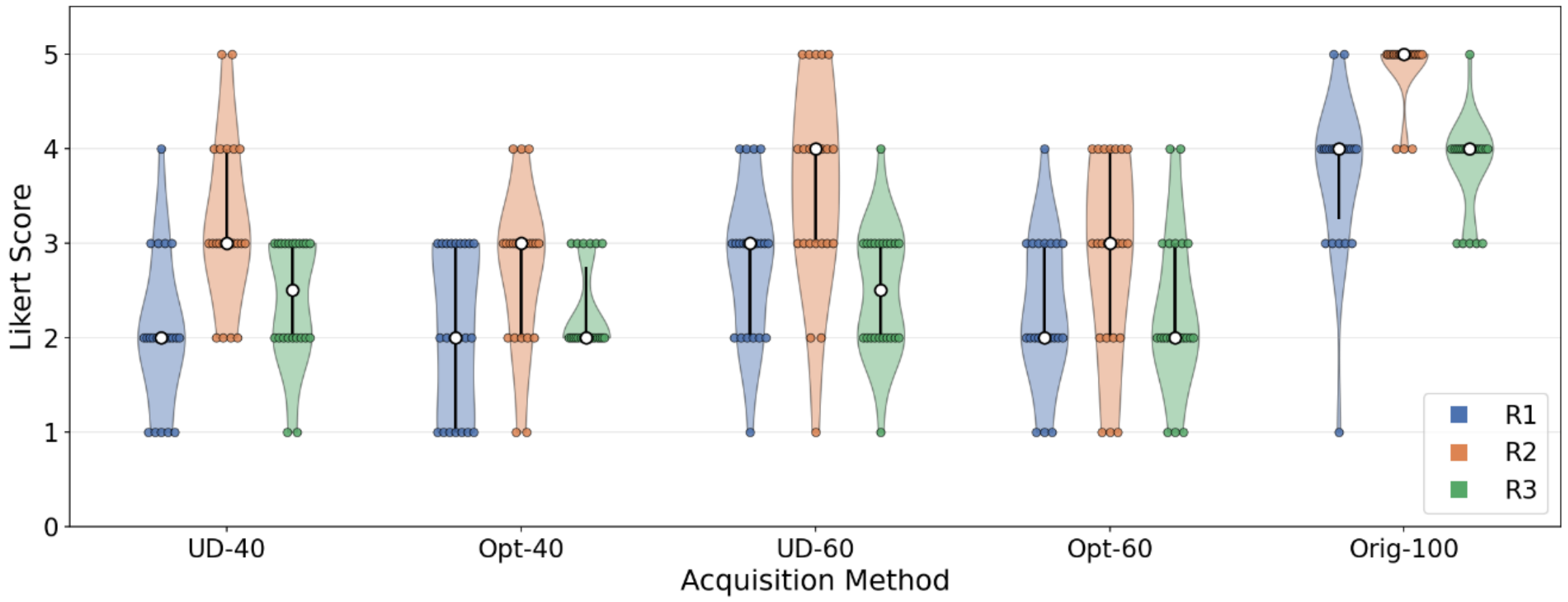

**Figure S3**: Violin plot summarizing the readers' image quality scores for each acquisition method in the pelvis. The white-filled circle denotes the median score, and the width and number of dots represent the number of responses for each image quality score.